\documentclass[aps,pra,twocolumn,amsmath,amssymb,superscriptaddress,longbibliography]{revtex4}

\usepackage{CJK}
\usepackage{graphicx} 
\usepackage{bm}
\usepackage[pdfstartview=FitH, CJKbookmarks=true, bookmarksnumbered=true, bookmarksopen=true, colorlinks=true, pdfborder=0 0 1, citecolor=blue, linkcolor=blue, linktocpage=true] {hyperref}
\usepackage{epstopdf} 

\begin{document}
\begin{CJK*}{GBK}{song}

\title{Topological states in disordered arrays of dielectric nanoparticles}

\author{Ling Lin}
\affiliation{Guangdong Provincial Key Laboratory of Quantum Metrology and Sensing $\&$ School of Physics and Astronomy,Sun Yat-Sen University (Zhuhai Campus), Zhuhai 519082, China}
\affiliation{State Key Laboratory of Optoelectronic Materials and Technologies, Sun Yat-Sen University (Guangzhou Campus), Guangzhou 510275, China}

\author{Sergey Kruk}
\affiliation{Nonlinear Physics Centre, Research School of Physics, Australian National University, Canberra ACT 2601, Australia}

\author{Yongguan Ke}
\affiliation{Guangdong Provincial Key Laboratory of Quantum Metrology and Sensing $\&$ School of Physics and Astronomy,Sun Yat-Sen University (Zhuhai Campus), Zhuhai 519082, China}
\affiliation{Nonlinear Physics Centre, Research School of Physics, Australian National University, Canberra ACT 2601, Australia}

\author{Chaohong Lee}
\email{lichaoh2@mail.sysu.edu.cn}
\affiliation{Guangdong Provincial Key Laboratory of Quantum Metrology and Sensing $\&$ School of Physics and Astronomy,Sun Yat-Sen University (Zhuhai Campus), Zhuhai 519082, China}
\affiliation{State Key Laboratory of Optoelectronic Materials and Technologies, Sun Yat-Sen University (Guangzhou Campus), Guangzhou 510275, China}

\author{Yuri Kivshar}
\email{yuri.kivshar@anu.edu.au}
\affiliation{Nonlinear Physics Centre, Research School of Physics, Australian National University, Canberra ACT 2601, Australia}

\date{\today}

%

\begin{abstract}
We study the interplay between disorder and topology for the localized edge states of light in topological zigzag arrays of resonant dielectric nanoparticles. 
We characterize topological properties by the winding number that depends on both zigzag angle and spacing between nanoparticles in the array. 
For equal-spacing arrays, the system may have two values of the winding number $\nu=0$ or $1$, and it demonstrates localization at the edges even in the presence of disorder, being consistent with experimental observations for finite-length nanodisk structures. 
For staggered-spacing arrays, the system possesses richer topological phases characterized by the winding numbers $\nu=0$, $1$ or $2$, which depend on the averaged zigzag angle and  disorder strength. 
In a sharp contrast to the equal-spacing zigzag arrays, staggered-spacing arrays reveal two types of topological phase transitions induced by the angle disorder, (i) $\nu  = 0 \leftrightarrow \nu  = 1$ and (ii) $\nu  = 1 \leftrightarrow \nu  = 2$. 
More importantly, the spectrum of staggered-spacing arrays may remain gapped even in the case of a strong disorder.
\end{abstract}

\maketitle

\section{Introduction}

Topological photonics emerged as a novel platform to realize robust optical circuitry protected against disorder~\citep{lu2016topological}.
The initial study of topological effects in photonics was inspired largely by direct analogies with similar effects in condensed matter physics for topologically nontrivial energy bands of electrons.
In photonics, the concept of topological phases provides non-traditional approaches in the search for innovative designs of advanced photonic devices as well as introduces novel physical effects and their applications~\citep{RevModPhys.91.015006}.
Spatially localized topological edge states have been predicted and realized in a variety of photonic systems, including topological lasers~\citep{st2017lasing}, single-photon transport~\citep{barik2018topological}, and topological solitons~\citep{mukherjee2020observation}.
The study of disorder in topological systems is very important because robustness against disorder is extensively employed as a measure of topologically protected features.
Anderson localization is a well-known effect of disorder in condensed-matter physics~\citep{PhysRev.109.1492}, which originates from interference of coherent waves, and it has been extended from electronic to other waves such as electromagnetic waves~\citep{riboli2014engineering,PhysRevB.50.9810}, acoustic waves \citep{hu2008localization}, and  matter waves \citep{billy2008direct, pasienski2010disordered}.
Strong spatial localization of light due to disorder can be realized in periodic or quasiperiodic dielectric structures, and it is represented by photonic band gaps.
For the systems with nontrivial topological phases, the interplay of topology-induced spatial localization (i.e. topological edge states) and disorder-induced Anderson localization becomes a very important problem, especially when disorder is no longer weak.
Here, for the first time to our knowledge, we address this important problem by studying the effects of disorder in topologically nontrivial zigzag arrays of coupled subwavelength optical resonators.

Many topological photonic systems which have been explored so far are based on waveguide geometries, and the size of their building blocks is therefore larger than the wavelength of light.
A zigzag array of coupled nanoresonators has emerged as the first nanoscale system with nontrivial topological states of light~\citep{poddubny2014topological}, and its topology-driven effects have been observed experimentally via lasing from excitonic-polaritonic nanoscale cavities~\citep{st2017lasing} and third-harmonic generation from nonlinear arrays of dielectric nanoresonators~\citep{kruk2019nonlinear}.

\begin{figure}[h]
  \includegraphics[width = 0.8 \columnwidth ]{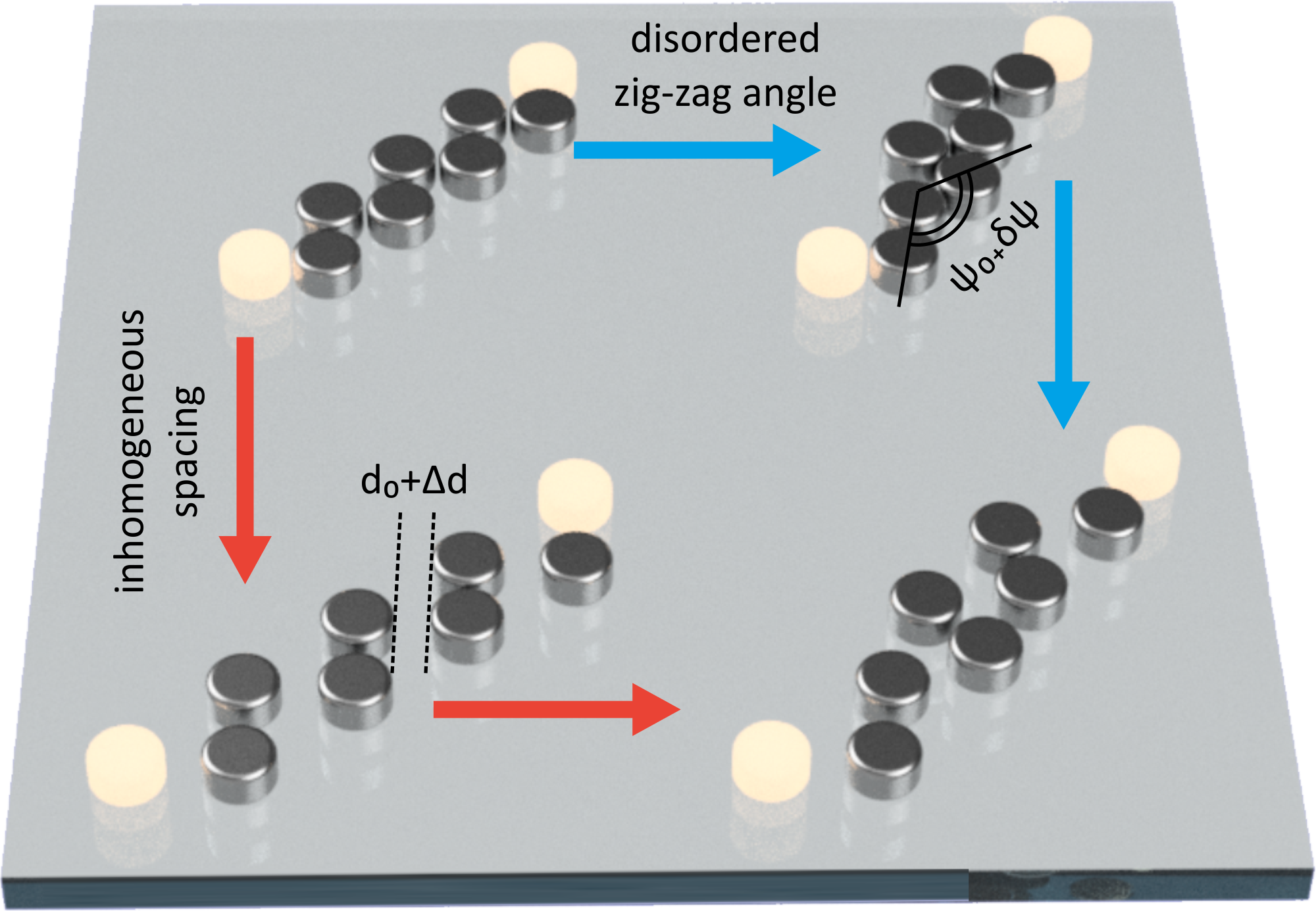}
  \caption{\label{fig:FIG_ConceptualFig}
 Schematic diagram of four different types of topologically nontrivial zigzag arrays of dielectric nanoparticles.
 The Mie-resonant nanodisks are arranged as equal spacing (top), staggered spacing (bottom), with no disorder in the zigzag angles (left)
 and with disorder in the zigzag angles (right).
  }
\end{figure}

In this paper, we study the interplay between disorder and topology in the zigzag arrays of resonant dielectric nanoparticles.
We consider four types of topological arrays with equidistant and non-equidistant spacing of particles, as illustrated in Fig.~\ref{fig:FIG_ConceptualFig}, and analyze their topological phases and disorder-induced topological phase transitions. 
We also present experimental results for Mie-resonant dielectric nanoparticles which confirm the basic predictions of the effect of disorder on topological properties, however being limited by the length of arrays.
Remarkably, for staggered-spacing arrays, we observe that the topological spectral gap can survive even in the presence of strong disorder, the effect that was never observed before for any topological system, to the best of our knowledge.

The paper is organized as follows.
In Sec.~\ref{Sec:equal_spacing}, we introduce our model of a zigzag array of dielectric nanoparticles.
In Sec.~\ref{SecWinding}, we study the topological properties of an ideal (regular) array and its disordered modifications.
In Sec.~\ref{Sec:Localization_length}, we use the zero-energy localization method to probe the disorder-induced topological transitions.
In Sec.~\ref{Sec:generalized_model}, we generalize our model of the topological zigzag array from the equal-spacing to staggered-spacing case.
Finally, Sec.~\ref{Sec:conclusion} concludes the paper.

\section{Equal-spacing zigzag arrays}
\label{Sec:equal_spacing}

\begin{figure}[h]
  \includegraphics[width = \columnwidth ]{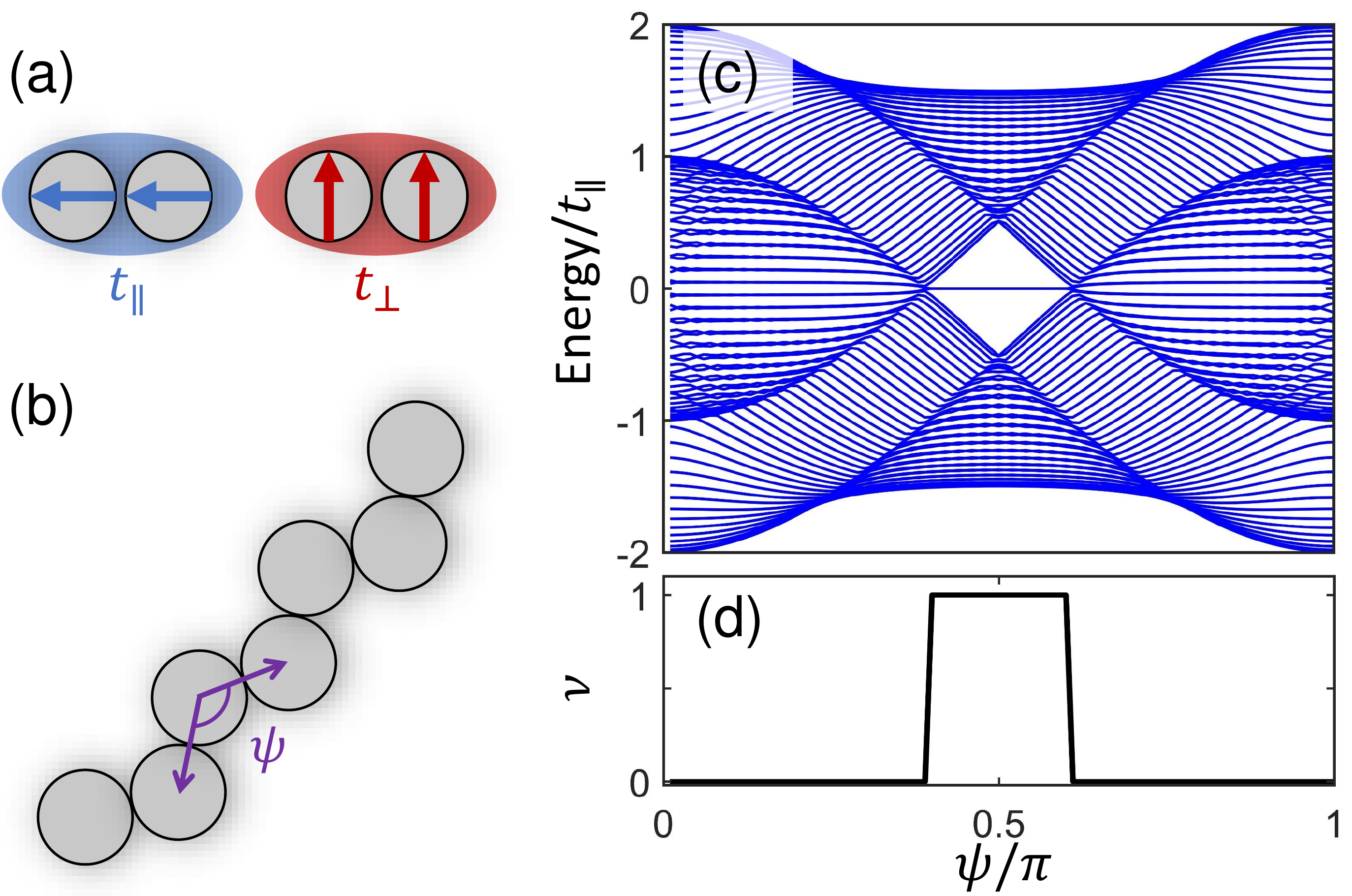}
  \caption{\label{fig:Zigzag_lattice}
 (a) Illustrations of polarization-dependent dipole interaction between nanoparticles.
 The arrows inside the particle indicate direction of electric or magnetic polarization.
 (b) Schematic illustration of the zigzag array.
 (c) Energy spectrum as a function of  zigzag angle $\psi$ under open boundary condition.
 (d) Winding number $\nu$ calculated from Eq.~\eqref{eqn:momentum_winding_number}.
  }
\end{figure}

It was established that topologically nontrivial states may appear in zigzag arrays of nanoparticles with polarization-dependent interaction between electromagnetic modes~\citep{Slobozhanyuk2015Subwavelength,small.201603190,kruk2019nonlinear, poddubny2014topological}.
Here, we follow those earlier predictions and consider the electric or magnetic polarizations of dielectric nanoparticle in the $x-y$ plane,
as shown in Fig.~\ref{fig:Zigzag_lattice}(a), coupled due to the dipole-dipole interaction. Such polarization-dependent interaction between two neighboring nanoparticles can be written in the form~\citep{small.201603190}
\begin{equation}
{V^{(j,j')}} = {t_\parallel }\vec e_\parallel ^{(j,j')} \otimes \vec e_\parallel ^{(j,j')} + {t_ \bot }\vec e_ \bot ^{(j,j')} \otimes \vec e_ \bot ^{(j,j')},
\end{equation}
where $\vec e_\parallel ^{(j,j')} $ and $\vec e_ \bot ^{(j,j')}$ denote respectively the identity vectors parallel and perpendicular to the link vector $\vec r_j - \vec r_{j'}$. $V^{(j,j')}=V^{(j',j)}$, and the coupling decays with the distance between the $j$th and $j'$th nanoparticles
due to the dipole-dipole interaction.
It is reasonable to keep only the nearest-neighboring interaction~\citep{Slobozhanyuk2015Subwavelength,small.201603190,kruk2019nonlinear, poddubny2014topological}.

By decomposing the polarization into $x$ and $y$ modes, the Hamiltonian of such zigzag arrays can be written as
\begin{equation}
\label{eqn:ZZ_Ham}
\hat H = \sum\limits_{j,\nu } {{E_0}a_{j,\nu }^\dag {a_{j\nu }}}  + \sum\limits_{ j ,\nu ,\nu '} {a_{j,\nu }^\dag V_{\nu ,\nu '}^{(j,j+1)}{a_{j+1,\nu '}}+h.c.},
\end{equation}
with
\begin{equation}
\label{eqn:tunneling_matrix}
V_{\nu ,\nu '}^{j,j+1} = \left\{ \begin{array}{l}
{t_\parallel }{\cos ^2}(\psi/2) + {t_ \bot }{\sin ^2}(\psi/2) ,\quad \nu  = \nu ' = x,\\
{t_\parallel }{\sin ^2}(\psi/2)  + {t_ \bot }{\cos ^2}(\psi/2) ,\quad \nu  = \nu ' = y,\\
\left( {{t_\parallel } - {t_ \bot }} \right)\sin (\psi/2) \cos (\psi/2) ,\quad \nu  \ne \nu'.
\end{array} \right.
\end{equation}
Here, $a_{j,\nu}^{\dag}$ and $a_{j,\nu}$ are the creation and annihilation operators for a $\nu$ mode at the $j$th nanoparticle,
 and $\psi$ is the zigzag angle, as indicated in Fig.~\ref{fig:Zigzag_lattice}(b).
$E_0$ is the on-site potential which causes an overall energy shift and hence is neglected.
The ratio $t_\parallel/t_\bot$ depends on the interaction, e.g., $t_\parallel/t_\bot=-2$ for the dipole-dipole interaction,
 and $t_\parallel/t_\bot = -4$ for the quadrupole-quadrupole interaction.
Here, we consider the case of $t_\parallel/t_\bot=-2$.

\section{Winding number}\label{SecWinding}
\subsection{Ideal arrays}
We start with the study of an ideal system without disorder.
In this case, an infinite system is translationally invariant.
Under periodic boundary condition, the Bloch Hamiltonian reads~\citep{Slobozhanyuk2015Subwavelength},
\begin{equation}
\label{eqn_Zizag_Hk}
H\left( K \right) = \left( {\begin{array}{*{20}{c}}
0&Q(K)\\
{{Q^\dag (K) }}&0
\end{array}} \right),
\end{equation}
with
\begin{equation}
Q(K) = {h_0}+ \vec h(K) \vec \sigma.
\end{equation}
Here, ${h_0} = t\left( {1 + {e^{ - iK}}} \right)$, and $\vec h(K) \equiv (h_x,h_y,h_z) = \left[ {\frac{\Delta }{2}{e^{ - iK}}\sin \psi ,\;0,\;\frac{\Delta }{2}\left( {1 + {e^{ - iK}}\cos \psi } \right)} \right]$ are components of an effective magnetic field with $t = \left( {{t_\parallel } + {t_ \bot }} \right)/2$ and $\Delta  = \left( {{t_\parallel } + {t_ \bot }} \right)$. $\vec \sigma\equiv (\sigma_x,\sigma_y,\sigma_z)$ are the Pauli matrices.

\begin{figure}[h]
  \includegraphics[width = \columnwidth ]{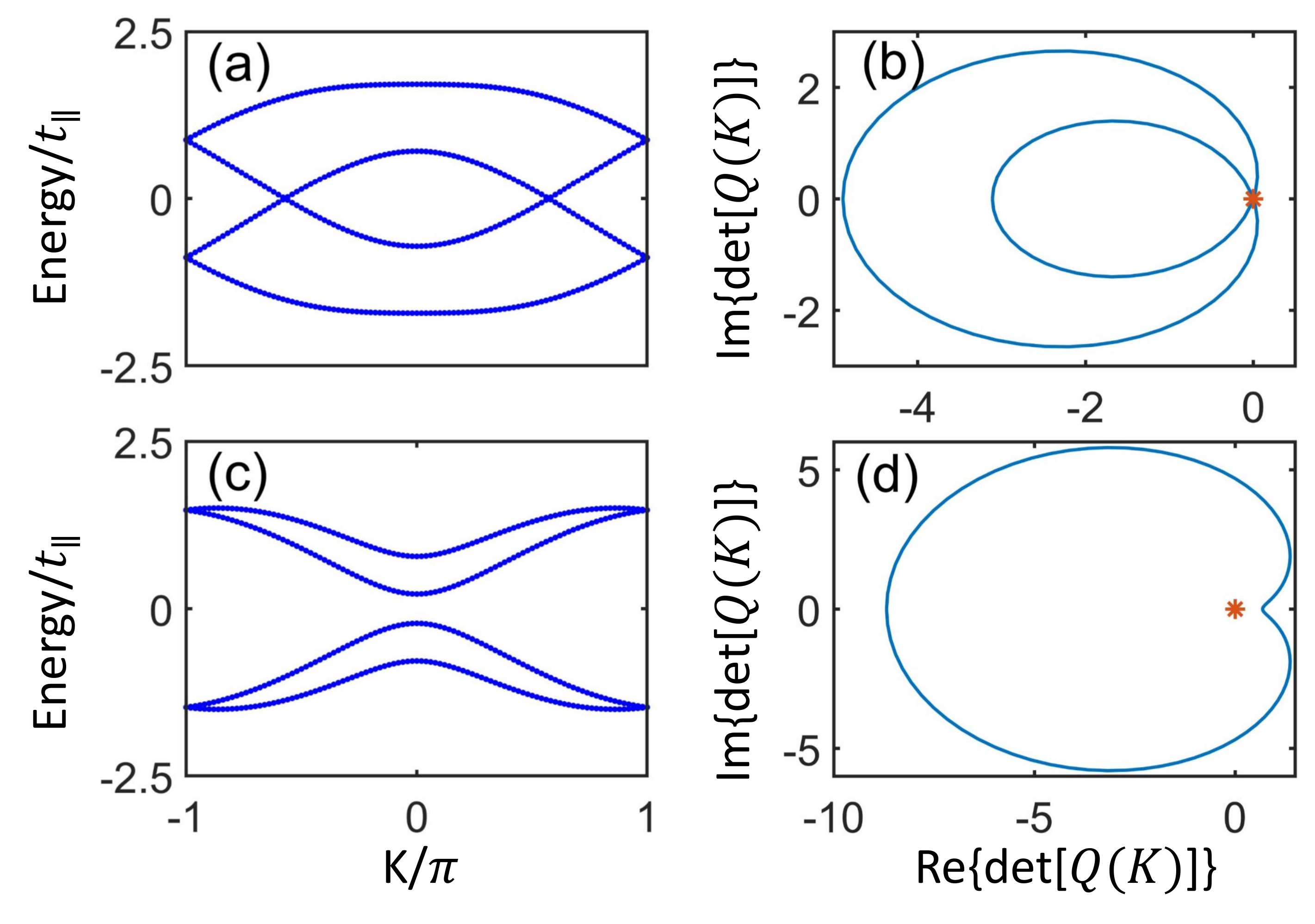}
  \caption{\label{fig:FIG_Spectrum_and_Winding}
 (a-d) Band structure (left column) and $\mathrm{det} [Q(K)]$ (right column) for different values of the zigzag angle $\psi$.
 (a-b) $\psi = 0.2\pi$: the system is gapless, and the curve of $\mathrm{det} [Q(K)]$ exactly lies on the origin.
 (c-d) $\psi = 0.44\pi$: the system is gapped, and the winding number is well defined.
  }
\end{figure}

According to the classification of topological insulators and superconductors, this system belongs to the so-called BDI class~\citep{ryu2010topological}, meaning a certain ``Cartan label" given to the corresponding symmetric space in \'Elie Cartan's classification scheme dating back to 1926.
The topological invariant of such systems is presented by the winding number,
\begin{eqnarray}
\label{eqn:momentum_winding_number}
\nu &=& \frac{1}{{2\pi i}}\int\limits_{ - \pi }^\pi  {dK~{{\rm{Tr}}\left[ {{Q^{ - 1}}\left( K \right){\partial _K}Q\left( K \right)} \right]} }  \nonumber \\
&=&   \frac{1}{{2\pi i}}\int\limits_{ - \pi }^\pi  {dK~\frac{{d\ln \left\{\det \left[Q\left( K \right)\right]\right\}}}{{dK}}}  \nonumber \\
&=& - \frac{1}{{2\pi }}\oint {d~\arg \left\{\det \left[Q\left( K \right)\right]\right\}}.
\end{eqnarray}
According to bulk-boundary correspondence, the nontrivial topological invariant in a bulk is always accompanied by the existence of
topological edge states at the boundaries. Under open boundary condition, the energy spectrum  with respect to the zigzag angle is shown in Fig.~\ref{fig:Zigzag_lattice}(c).
The spectrum can be either gapless or gapped, where in-gap zero-energy states appear only in the gapped case, implying topologically nontrivial properties.
To confirm this, we calculate the winding number as a function of the zigzag angle; see Figs.~\ref{fig:Zigzag_lattice}(d-e), respectively.
Indeed, in the gapless case the winding number vanishes, while in the gapped case the winding number equals to $1$.
Our results are consistent with earlier studies~\citep{Slobozhanyuk2015Subwavelength}, where
topologically nontrivial phase was predicted for $|\psi  - \pi /2| < \psi_{thre}$ with the threshold value $\psi_{thre}=\arcsin |\left( {{t_\parallel } + {t_ \bot }} \right)/\left( {{t_\parallel } - {t_ \bot }} \right)| /2$.

We notice here that the original winding number~\eqref{eqn:momentum_winding_number} is ill-defined in the gapless case.
When the energy bands are gapless, the trajectory of ${ \det Q}$ in the complex plane crosses the original point where its angle is ambiguous; see Figs.~\ref{fig:FIG_Spectrum_and_Winding}(a-b).
However, a tiny perturbation without breaking the chiral symmetry does not change the topological property.
To fix this problem, in numerical calculations we need shifting $\mathrm{det} Q(K)$ and redefine $\nu' = \nu \; \mathrm{mod}\;2$ to avoid this ambiguity~\citep{Slobozhanyuk2015Subwavelength}.
For completeness, we also show the energy bands and the trajectory in the complex plane for the gapped case; see Figs.~\ref{fig:FIG_Spectrum_and_Winding}(c-d).
${\det Q}$ is always nonvanishing, and the winding number is $1$.

\subsection{Disordered arrays}
Next, we study the effects of disorder on the zigzag array.
When the zigzag angle between nanoparticles experiences random variations, we may expect the well-known Anderson localization to occur.
At the same time, in the topologically nontrivial case characterized by the winding number $\nu=1$, we expect topological edge states, which are localize exponentially at one or two edges.
Since the zigzag-angle disorder can be realized in experiment~\citep{kruk2019nonlinear}, we study the interplay between topology-induced edge localization and the Anderson localization.
In the following, we keep $t_\parallel = 2, t_\bot = -1$ fixed and assume that disorder is applied to the zigzag angle,
$\psi=\psi_0 + \delta \psi_i$, with averaged zigzag angle $\psi_0$ and $\delta \psi_i = W\epsilon_i$, where $\epsilon_i$ is a random number distributed uniformly within $[-\pi/2, \pi/2]$.
Such disorder affects only the coupling, and it will not break the chiral symmetry.
As the range of zigzag angles $\psi$ is limited, see Fig.~\ref{fig:FIG_LocPhaseDiagram}(a), the strength of random angles $\delta \psi_i $ should be limited within $\delta \psi \in [-\pi/2, \pi/2] $.
Thus, the strength of disorder in this context is restricted to the domain $W \in [0, 1]$.

The winding number~\eqref{eqn:momentum_winding_number} is obtained generally in the translation-invariant systems, and usually the straighforward definition fails for disordered systems.
If disorder does no break the essential discrete symmetry, i.e. the chiral symmetry, the winding number can be still extracted via a real-space formula~\citep{PhysRevLett.113.046802, PRODAN20161150}.
The calculation of real-space winding number relies on the homotopically equivalent flat band Hamiltonian $\hat Q=\hat P_+ - \hat P_-$, in which
\begin{equation}
\left\{ \begin{array}{l}
{{\hat P}_ + } = \sum\limits_{n \in \{ {E_n} > 0\} } {|{\varphi _n}\rangle \langle {\varphi _n}|} \\
{{\hat P}_ - } = \sum\limits_{n \in \{ {E_n} < 0\} } {|{\varphi _n}\rangle \langle {\varphi _n}|}
\end{array} \right.
\end{equation}
are the projector onto the subspace with positive or negative energies, respectively.
It can be found that $\hat Q$ shares the same eigenstates with Hamiltonian $\hat H$~\citep{ryu2010topological,RevModPhys.88.035005}.
For chiral-symmetric systems, $Q$ can be composed by $Q = Q_{AB} + Q_{BA}$, with  ${Q_{AB}} = {\Gamma _A}Q{\Gamma _B}$, $Q_{BA} = {\Gamma _B}Q{\Gamma _A}$.
Here $\Gamma_{A}$ and $\Gamma_B$ are respectively the projectors onto the two sublattices, labelled by $A$ and $B$.
Then, the winding number $\nu$ can be calculated in real space via the formula~\citep{PhysRevLett.113.046802, Science362,prodan2016bulk}
\begin{equation}
\label{eqn:winding_number}
\nu  =  - {\rm{Tr}}\left\{ {{Q_{ BA }}\left[ {X,{Q_{AB }}} \right]} \right\}
\end{equation}
where $X$ is the position operator.
This formula remains valid even in the presence of disorder~\citep{PhysRevLett.113.046802, PRODAN20161150,schulz2016topological,Science362}, and it has been used widely in the studies of disorder and topology \citep{PhysRevB.90.125143, PhysRevB.100.205302,PhysRevB.101.224201,Maffei2018,PhysRevLett.122.126801}.

\begin{figure}[htp]
  \includegraphics[width = \columnwidth ]{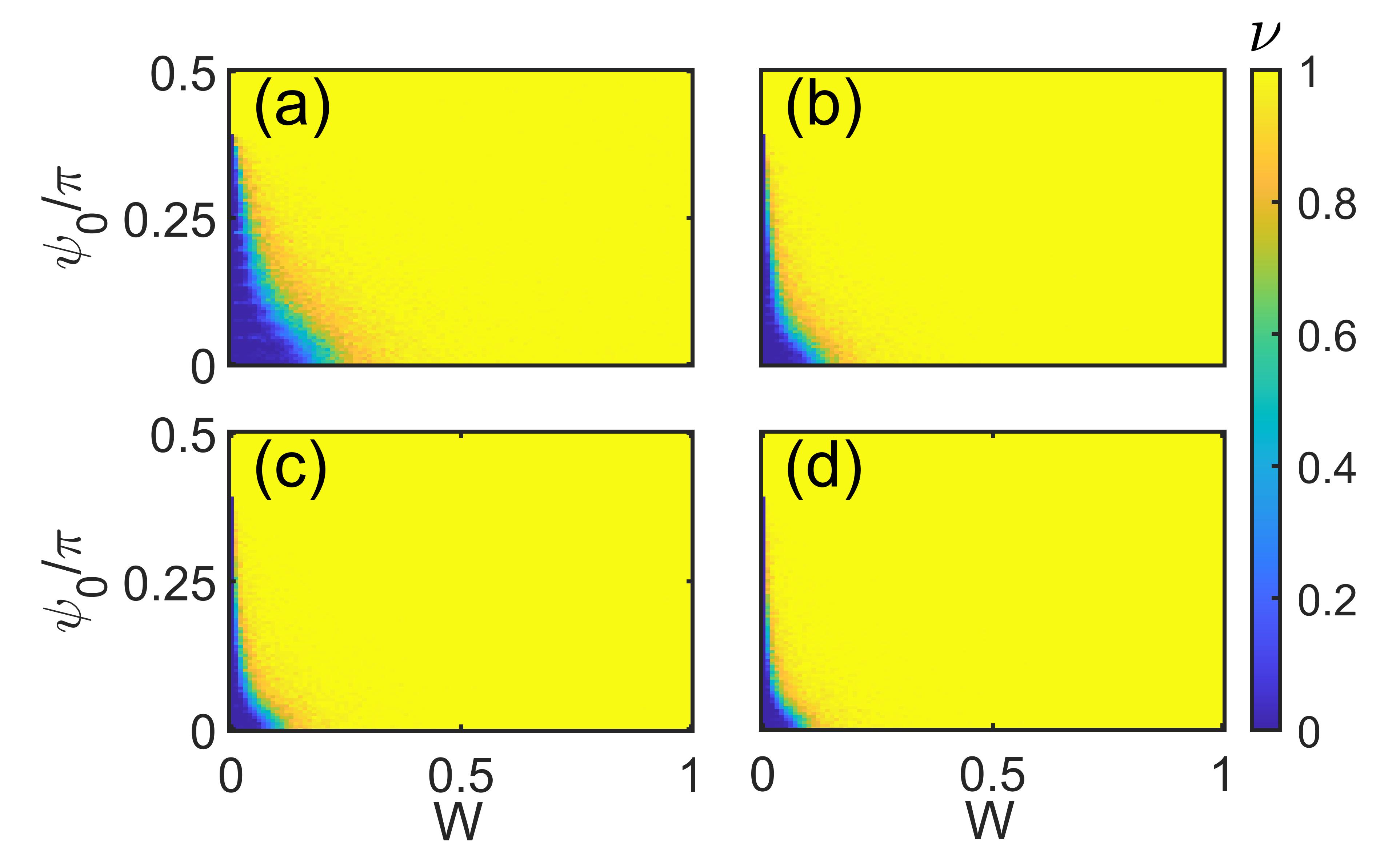}
  \caption{\label{fig:FIG_Scaling_WN}
 Winding number for different zigzag angles $\psi_0$ and strength of disorder $W$.
The cases (a-d) are calculated for $62,\;122,\;162$ and $202$ nanoparticles, respectively.
 }
\end{figure}

In our system, the gapless regime results in ambiguity when we calculate the winding number.
However, when the disorder is present, this quantity is shown to be valid even if the gap is closed \citep{PhysRevLett.113.046802,PhysRevB.89.224203}.
We calculate the phase diagram of the winding number in the parameter space $(\psi_0, W)$; see Fig.~\ref{fig:FIG_Scaling_WN}.
As turns out, the topology of such a system is nontrivial for large area where $W>0$.
The appearance of a small area of the trivial phase can be understood as the finite-size effect, and it will tend to disappear in the thermodynamic limit $L\rightarrow \infty$.
This analysis suggests that the system may become topological when disorder of the zigzag angle is added.
In the regular case ($W=0$), the gapless region $|\psi_0  - \pi /2| > \psi_{thre}$ is topologically trivial.
Once the disorder is added, it immediately changes to topologically nontrivial.
Hence, the case $W=0$ can be understood as a critical point of the phase transition for the region $|\psi_0  - \pi /2| > \psi_{thre}$.
When disorder is present, the system is pushed towards a topological phase.

\section{Zero-energy localization length}
\label{Sec:Localization_length}

Above, we have found that our system is topologically nontrivial when the angle disorder is included.
We notice that, for one-dimensional chiral-symmetric systems, a topological transition is usually accompanied by a divergence of the localization length at $E=0$~\citep{PhysRevLett.113.046802, PhysRevLett.106.057001, Fulga2011Scattering}.
The Lyapunov exponent, which is related to the inverse of the localization length, will vanish at the point of the topological transition
and then it changes its sign.
This is an important quantity to reveal the phase boundary.

\subsection{Transfer matrix approach}
First, we start with the transfer matrix method and employ it for our system.
According to the theory of Anderson localization~\citep{PhysRev.109.1492,RevModPhys.80.1355}, a one-dimensional lattice system becomes an insulator as long as any even weak disorder appears, and the electron wavefunction decays exponentially characterizing a localized state,
\begin{equation}
|\varphi \left( r \right) |\sim \exp \left( { - \frac{{|r - {r_0}|}}{\xi }} \right) ,
\end{equation}
where $\xi$ is called the \emph{localization length}.
By employing the transfer-matrix technique \citep{PhysRevLett.70.982,RevModPhys.69.731}, we calculate the localization length of our one-dimensional system.
The discrete Schrodinger equation for the eigenstates of our model reads
\begin{eqnarray*}
\label{eqn:discrete_S_eqn}
 t_x^{i,i + 1}\varphi _i^x + t_x^{i + 1,i + 2}\varphi _{i + 2}^x + t_{xy}^{i,i + 1}\varphi _i^y+ t_{xy}^{i + 1,i + 2}\varphi _{i + 2}^y = E\varphi _{i + 1}^x \\
 t_y^{i,i + 1}\varphi _i^y + t_y^{i + 1,i + 2}\varphi _{i + 2}^y + t_{xy}^{i,i + 1}\varphi _i^x + t_{xy}^{i + 1,i + 2}\varphi _{i + 2}^x = E\varphi _{i + 1}^y
\end{eqnarray*}
where $\varphi^x$ and $\varphi^y$ are the wavefunctions of the $x$ and $y$ polarized modes at $i$th site, respectively.
$t_\sigma^{i,i+1} \; (\sigma=x, y)$ is the coupling strength between $i$-th and $(i+1)$-th particles,
 \begin{equation}
\label{eqn:tunneling_matrix_txtytxy}
\left\{ \begin{array}{l}
t_x \equiv {t_\parallel }{\cos ^2}\psi/2  + {t_ \bot }{\sin ^2}\psi/2,\\
t_y \equiv{t_\parallel }{\sin ^2}\psi/2  + {t_ \bot }{\cos ^2}\psi/2 ,\\
t_{xy} \equiv\left( {{t_\parallel } - {t_ \bot }} \right)\sin \psi/2 \cos \psi/2.
\end{array} \right.
\end{equation}
Equation~\eqref{eqn:discrete_S_eqn} can be written in a matrix form
\begin{equation}
{\Phi _{i + 1}} = M_i{\Phi _i},\;\;{\Phi _i} =
\left( {\begin{array}{*{20}{c}}
{\varphi _i^x}\\
{\varphi _i^y}\\
{ - t_x^{i,i + 1}\varphi _{i + 1}^x - t_{xy}^{i,i + 1}\varphi _{i + 1}^y}\\
{ - t_{xy}^{i,i + 1}\varphi _{i + 1}^x - t_y^{i,i + 1}\varphi _{i + 1}^y}
\end{array}} \right)
\end{equation}
where the matrix $M_i$,
 \begin{equation}
 {M_i} = \left( {\begin{array}{*{20}{c}}
0&0&{{t_y}/{t_\parallel }{t_ \bot }}&{ - {t_{xy}}/{t_\parallel }{t_ \bot }}\\
0&0&{ - {t_{xy}}/{t_\parallel }{t_ \bot }}&{{t_x}/{t_\parallel }{t_ \bot }}\\
{{t_x}}&{{t_{xy}}}&{E{t_y}/{t_\parallel }{t_ \bot }}&{ - E{t_{xy}}/{t_\parallel }{t_ \bot }}\\
{{t_{xy}}}&{{t_y}}&{ - E{t_{xy}}/{t_\parallel }{t_ \bot }}&{E{t_x}/{t_\parallel }{t_ \bot }},
\end{array}} \right)
 \end{equation}
is the so-called transfer matrix between $i$-th and $(i+1)$-th sites; we use the fact that $t_x t_y-t_{xy}^2={t_\parallel }{t_ \bot }$),
 and also apply a constraint $\mathrm{det}(M_i)=1$.
Thus, we are able to know the $L$-th wave function of the entire system by multiplying iteratively the transfer matrices,
\begin{equation}
{\Phi _L} = \prod\limits_{i = 1}^{L - 1} {{M_i}} {\Phi _1}.
\end{equation}
The Lyapunov exponent $\lambda$, that is the inverse of localization length $\lambda \equiv 1/\xi$, can be calculated
from the eigenvalue problem,
\begin{equation}
\mathcal{M} = \left[ {\mathop {\lim }\limits_{L \to \infty } \left( {\prod\limits_{i = 1}^L {{M_i}} } \right){{\left( {\prod\limits_{i = 1}^L {{M_i}} } \right)}^\dag }} \right].
\end{equation}
Here, we employ the efficient numerical method of Ref.~\citep{Mackinnon1983The,PhysRevLett.47.1546} to calculate the localization length.

\begin{figure*}[htp]
  \includegraphics[width = \textwidth ]{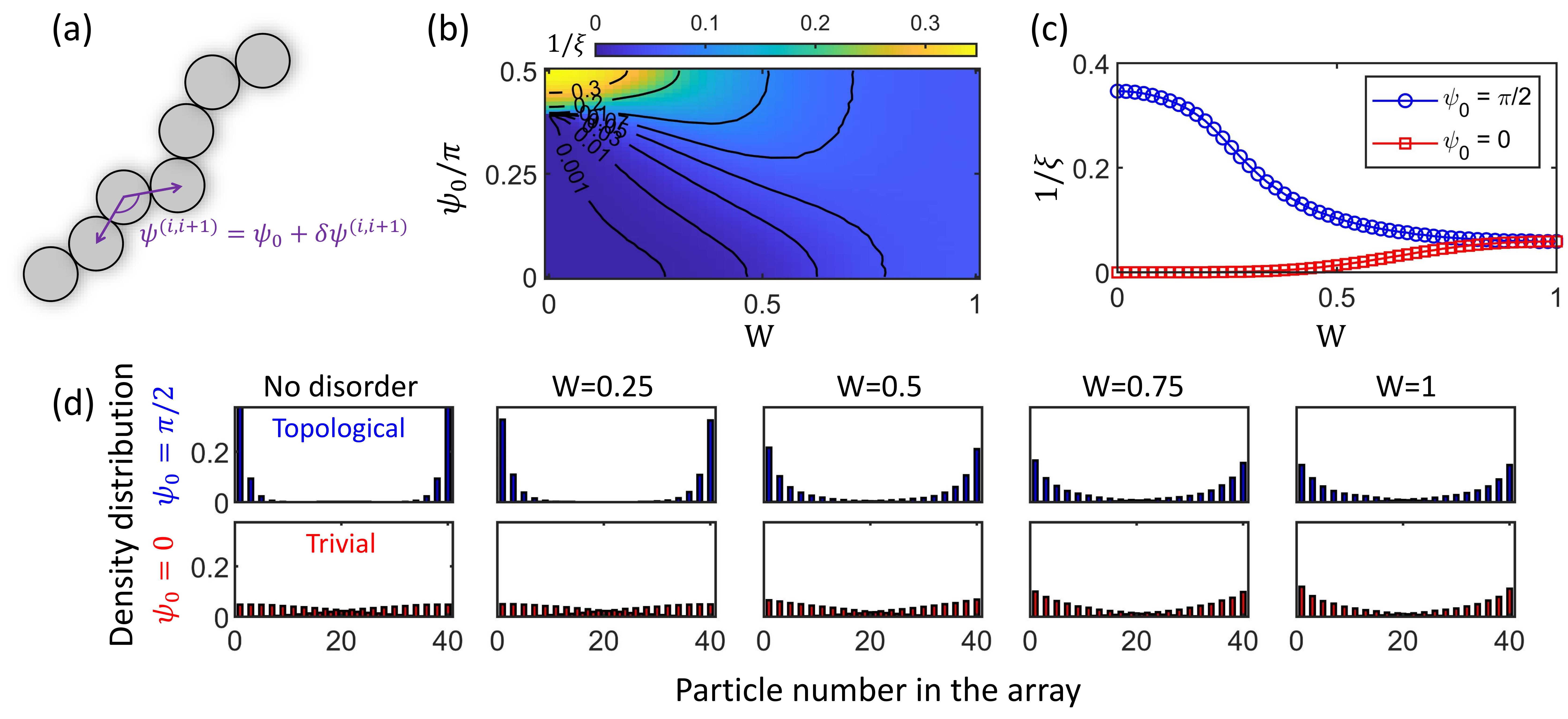}
  \caption{\label{fig:FIG_LocPhaseDiagram}
 (a) Schematic of disordered zigzag array.
 (b) Inverse localization length $1/\xi$, obtained from transfer matrix method, versus the disorder strength $W$ and the zigzag angle $\psi_0$.
 Black lines are the contours of $1/\xi$.
 Due to the symmetry about $\psi_0=\pi/2$, we only show the results of $\psi_0 \in [0, \pi/2]$.
 (c) The inverse localization length of $\psi_0 = \pi/2, \;0$ vs. disorder strength $W$.
 (d) Density distributions of the eigenstates which are \emph{the closest to $E=0$} for different disorder strengths.
 Upper and lower panels correspond to $\psi_0 = \pi/2$ and $0$, respectively.
 In our calculations, we choose 40 particles and average over 500 random realizations.
 The density of each particle is the summation of squared absolute value of the $x$- and $y$-mode wavefunction.
 }
\end{figure*}

\begin{figure*}[!htp]
	\includegraphics[width = \textwidth ]{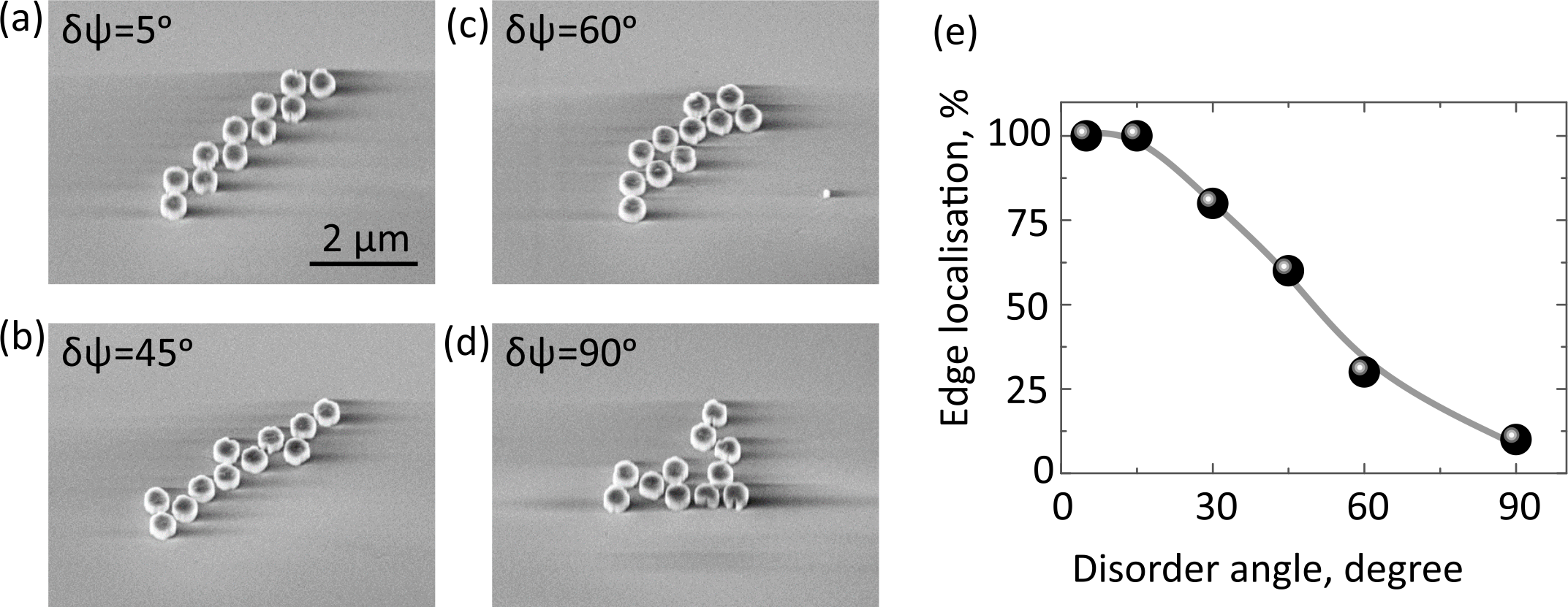}
	\caption{\label{fig:FIG_Exp}
		Experimental results for disordered nanoparticles arrays.
		(a)-(d) Scanning electron microscopy (SEM) image of fabricated zigzag arrays of Mie-resonant dielectric nanodisks with disordered strengths $\delta \psi=5^{\circ},\ 45^{\circ},\ 60^{\circ},\ 90^{\circ}$, respectively.
		The averaged zigzag angle is $\psi_0=\pi/2$, which is the same setup as Ref.~\citep{kruk2019nonlinear}.
		(e) Edge localization ratio extracted from experimental observations as a function of disorder angle.
	}
\end{figure*}

\subsection{Topology and localization length}
In Fig.~\ref{fig:FIG_LocPhaseDiagram}(b), we show the phase diagram of the inverse localization length $1/\xi$  with respect to $(W,\psi_0)$ at $E=0$
In the gapped case, the zero-energy localization length is exactly the length of localization of the edge state.
Disorder will increase the localization length; see blue curve in Fig.~\ref{fig:FIG_LocPhaseDiagram}(c).
In the gapless case, the zero-energy localization length diverges, and disorder will result in a decrease of the localization length;
see the red curve in Fig.~\ref{fig:FIG_LocPhaseDiagram}(c).
We find that divergence of the localization length occurs only in the gapless region $|\psi_0  - \pi /2| > \psi_{thre}$ when $W=0$, and we do not find any divergence of the localization length when $W>0$.
This implies that the system remains in the same topological phase apart from the region $(|\psi_0  - \pi /2| > \psi_{thre},W=0)$.
Thus, we may conclude that the topologically nontrivial system remains topological when $W>0$.
As we find the winding number of the system remains nontrivial for $W>0$, the system should present topologically nontrivial phenomena.
One of the features in one-dimensional topological systems is the appearance of topological edge states.
In Fig.~\ref{fig:FIG_LocPhaseDiagram}(d), we show the density distribution of the eigenstates with $E \approx 0$ under different strengths of disorder.
We find that the topological edge states with averaged zigzag angle $\psi_0 = \pi/2$ are preserved as the value of $W$ grows.
Meanwhile, with an increase of disorder strength, there appear localized states for the zigzag angle $\psi_0 = 0$.
For a weak disorder, the localization length of the edge state is so large that the edge state is difficult to distinguish for a small number of nanoparticles.
However, the edge-localization effect will become distinct when the system size is increased.

\subsection{Experimental results}
\label{Sec:subSec_Experimental_results}
We have verified our theoretical predictions in experiment.
For experiment, we fabricate many different types of disordered zigzag arrays of nanodisks placed on a glass substrate, as shown in Figs.~\ref{fig:FIG_Exp}(a-d).
The samples are obtained with the same platform and techniques as described earlier in Ref.~\citep{kruk2019nonlinear}.
To map topological states, we excite zigzag arrays by femtosecond laser pulses, and observe the generation of a third-harmonic signal, being a tool for mapping strong spatially-localization distributions of light in the arrays.
We set different strengths of the angle disorder, as shown in the examples of Figs.~\ref{fig:FIG_Exp}(a-d), and also in Supplementary Fig.~\ref{fig:FIG_Exp_Suppl}.
We analyze the edge localization
via the third-harmonic field in the zigzag arrays composed of eleven nanodisks with the averaged zigzag angle $\psi_0=\pi/2$
and different angle disorder.
We find that the number of distinguishable edge states decreases when the disorder strength increases, see
Fig.~\ref{fig:FIG_LocPhaseDiagram}(e), however some of the edge is clearly observed for most of the cases.
The reduction of the edge localization observed in experiment corresponds directly to the effect predicted numerically, and it can be understood as follows.
The localization length increases with the strength of disorder $W$ for the averaged zigzag angle $\psi_0=\pi/2$,
see Fig.~\ref{fig:FIG_LocPhaseDiagram}(d).
When the localization length of the edge state excesses the length of the array, we are not able to distinguish it from other states,
similarly observed for the trivial case when $\psi_0=0$, see Fig.~\ref{fig:FIG_LocPhaseDiagram}(d).
For example, the localization length of the edge state is about $\xi=10$ at $W=0.5$ (this corresponds to the disorder angle $\delta \psi \approx \pi/4$).
This is almost the same length as the extension of our eleven-disks nanoparticle arrays employed in experiment.
Thus, while in general we confirm the basic prediction of the theory about the character of the topology-induced localization and its interplay with the disorder, it becomes difficult to recognize the edge states and their localization for stronger disorder when $W>0.5$ with finite-extent arrays.

\begin{figure}[htp]
  \includegraphics[width = \columnwidth ]{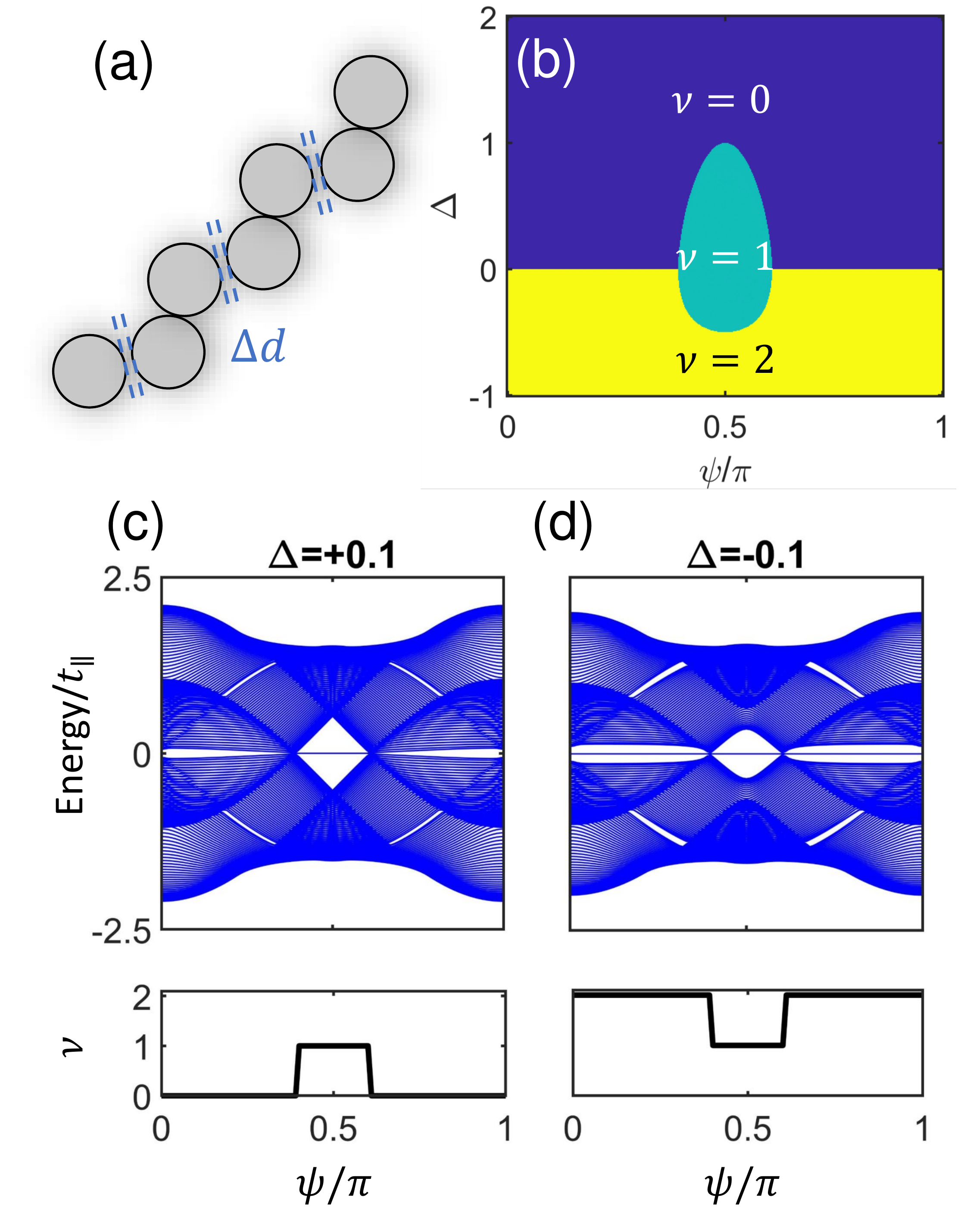}
  \caption{\label{fig:FIG_Spectrum_and_Winding_RE}
 (a) Schematic of a staggered-spacing zigzag array.
 $\Delta d$ indicates a larger (or less) space at the odd bond.
 (b) Phase diagram of winding number $\nu$ with respect to the bias $\Delta$ and the zigzag angle $\psi$.
 (c-d) Two typical energy spectra under open boundary condition and the corresponding winding number versus the zigzag angle.
 The winding number is determined from Eq.~\eqref{eqn:momentum_winding_number}.
 }
\end{figure}
\begin{figure*}[htp]
  \includegraphics[width = \textwidth]{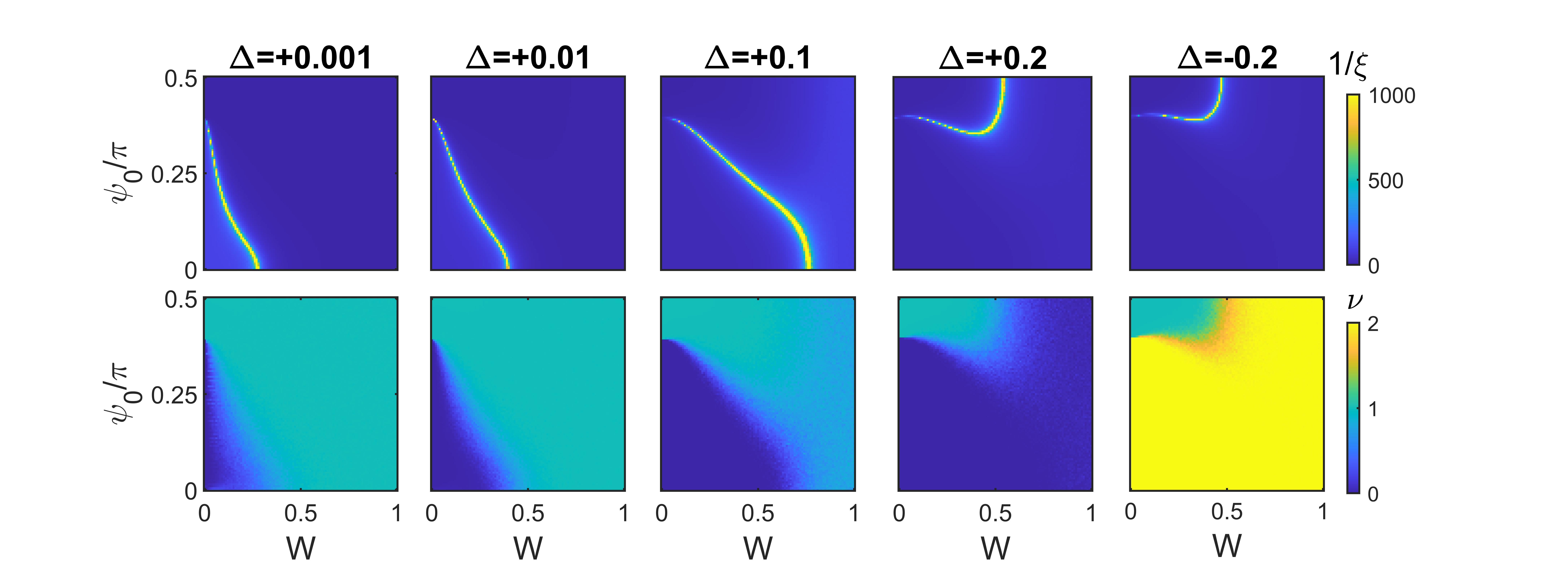}
  \caption{\label{fig:FIG_PhaseDiagram_NewDelta}
 \emph{Top row}: Localization length.
 Calculation of the transfer matrix iterated for at least $10^6$ times.
 \emph{Bottom row}: Winding number.
 Changes of the topological transition boundary for $\Delta<0$ is similar to $\Delta$, and thus we have omitted other cases for simplicity.
 In our calculations, we use 202 particles under periodic boundary condition, and average over 200 random realizations.
 }
\end{figure*}

\section{Staggered-coupling zigzag arrays}
\label{Sec:generalized_model}
Above, we have found theoretically that disorder-induced topological transitions may happen only at the gapless case when $|\psi_0  - \pi /2| > \psi_{thre}$.
Here, we consider a generalized model and explore richer topological phases and predict disorder-induced topological phase transitions,
for the first time to our knowledge.
We notice that the coupling between particles strongly depends on the distance between nanoparticles~\citep{poddubny2014topological}, and
in our model discussed above all nanoparticles are equally spaced.
In this section, we consider a staggered-spacing array in which the relative distance between nanoparticles alternates for odd and even bonds, as shown in Fig.~ \ref{fig:FIG_Spectrum_and_Winding_RE}(a).
In this case, the generalized Hamiltonian reads
\begin{equation}
{\hat H_2} = \sum\limits_{ j ,\nu ,\nu '} {\left( {1 + {\Delta ^{(j)}}} \right)a_{j,\nu }^\dag V_{\nu ,\nu '}^{(j,j+1)}{a_{j+1,\nu '}}} +h.c.,
\end{equation}
where $\Delta^{(j)}=0$ for even bond, and $\Delta^{(j)}=\Delta$ for odd bond.

\subsection{Ideal arrays}
We start with an ideal array when disorder is absent.
Chiral symmetry is still preserved in the presence of the extra term, and the topological property is still characterized by the winding number Eq.~\eqref{eqn:momentum_winding_number}.
The phase diagram of the winding number is obtained by varying the averaged zigzag angle and disorder strength; see Fig.~\ref{fig:FIG_Spectrum_and_Winding_RE}(b).
We uncover three topologically different phases characterized by the winding numbers $\nu=0,\; 1,\;2$.
When $\Delta = 0$, the generalized model becomes the original model of the zigzag array discussed above.
In Figs.~\ref{fig:FIG_Spectrum_and_Winding_RE}(c-d), we show two typical energy spectra under open boundary condition
and the corresponding winding numbers.
%
%
When there is no zero-energy edge mode, the topological phase is trivial, and therefore $\nu=0$.
When zero-energy edge modes appear, the zero-energy edge states are degenerate, and their numbers $n_{edge}=2,\; 4$
correspond to $\nu=1,\;2$, respectively.
This reflects the well-known bulk-edge correspondence.

\begin{figure}[!htp]
	\includegraphics[width = \columnwidth ]{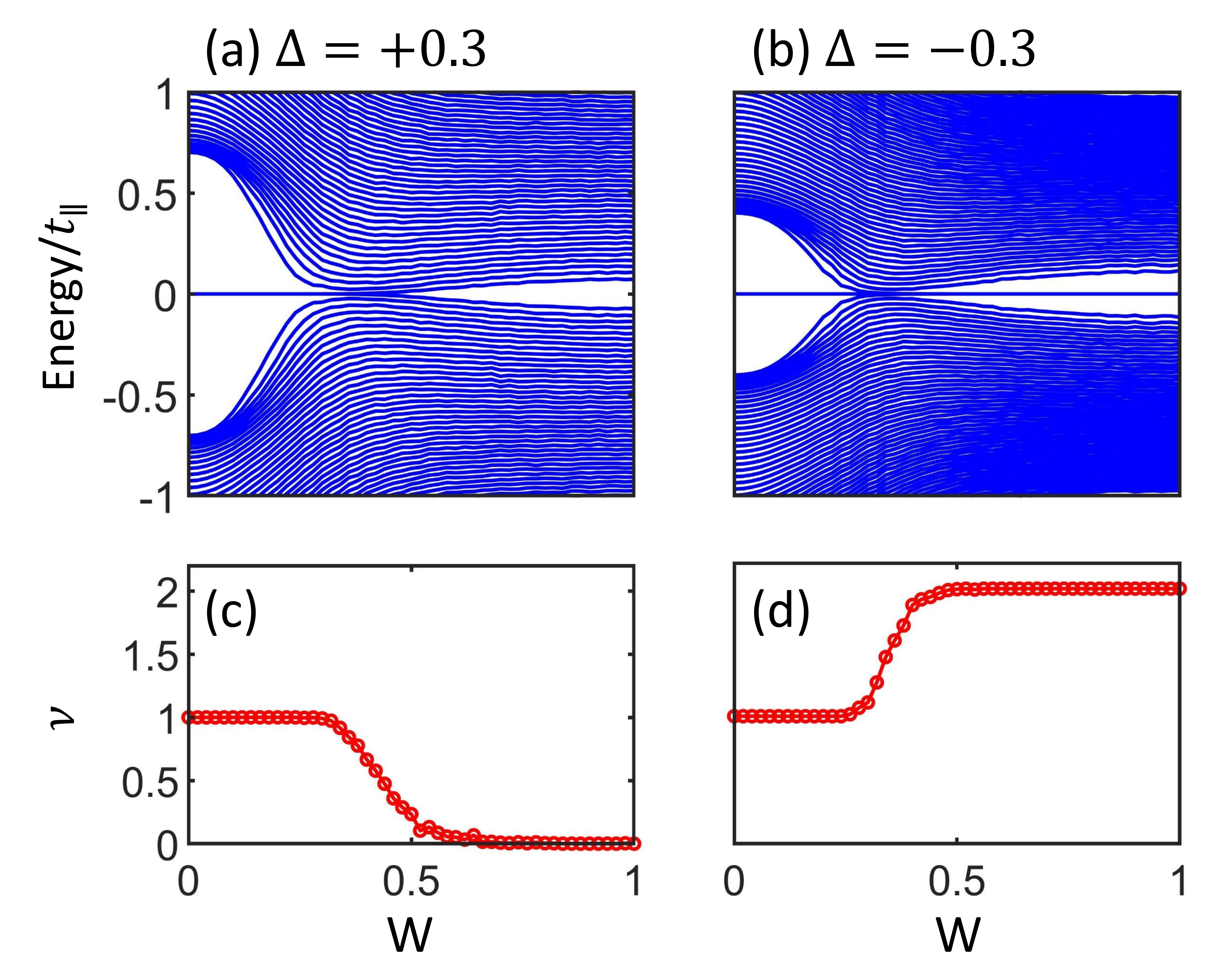}
	\caption{\label{fig:FIG_ScanRandomness}
		(a-b) Energy spectrum with respect to the strength of disorder $W$ near $E=0$ with $\Delta = \pm0.3$, respectively.
		Here, the averaged zigzag angle is $\psi_0=\pi/2$.
		Zero-energy edge modes in (a) are two-fold degenerate.
		Zero-energy edge modes in (b) are two-fold degenerate in the clean limit $W=0$, and then turn into four-fold degenerate after the phase transition.
		 (c) and (d) are the winding number of (a) and (b) respectively.
 		All results are averaged over 200 random realizations.
	}
\end{figure}

\subsection{Disordered arrays}
Next, we study the effects of disorder in zigzag angles.
In Fig.~\ref{fig:FIG_PhaseDiagram_NewDelta}, we show the phase diagram of the localization length in the parameter space $(\psi_0,\ W)$ as $\Delta$ changes (see the top row).
For each fixed $\Delta$, we find a narrow yellow line in the diagram with an extremely large localization length, indicating a boundary of a topological phase transition.
We calculate also the phase diagram of the winding number using Eq.~\eqref{eqn:winding_number}, as shown in the bottom row in Fig.~\ref{fig:FIG_PhaseDiagram_NewDelta}.
The boundary of the phase transition coincides with the one shown in the diagram of the localization length.
Besides, we notice that the region of trivial winding numbers shrinks to $W=0$ when $\Delta$ vanishes.
This is consistent with the case of $\Delta=0$ discussed above, where the system is topologically nontrivial in the presence of angle
disorder for any $W>0$.

As stated above, we observe three topologically distinct phases characterized by $\nu=0,\; 1,\; 2$.
Besides the disorder-induced topological transition between $\nu=0$ and $\nu = 1$, there is also the transition
between $\nu=1$ and $\nu=2$ when $\Delta < 0$.
The phase diagram of $\Delta < 0$ is similar to that of $\Delta > 0$, see the last column in Fig.~\ref{fig:FIG_PhaseDiagram_NewDelta}.
Next, we compare two examples of $\Delta  >0$ and $\Delta < 0$ in Fig.~\ref{fig:FIG_ScanRandomness}.
We identify a disappearance of two-fold degenerate edge states in Fig.~\ref{fig:FIG_ScanRandomness}(a), corresponding to the transition from $\nu=1$ to $\nu=0$.
In Fig.~\ref{fig:FIG_ScanRandomness}(b), we find that two-fold degenerate edge states are transformed into four-fold degenerate edge states, which characterises the transition from $\nu=1$ to $\nu=2$.
A change in the number of zero-energy edge states is a result of a topological transition, and the number of zero-energy edge states coincides with the winding number in Figs.~\ref{fig:FIG_ScanRandomness}(c-d).

When the deviation $|\Delta|$ is small, the spectral gap will be closed when disorder becomes stronger.  Intuitively, this happens because the disorder becomes dominating, and it "wipes away" the details of the staggered-spacing distribution.
However, in both these cases of Fig.~\ref{fig:FIG_ScanRandomness}, we observe that the spectral gap does not vanish for a strong disorder.
This result differs from a common belief that stronger disorder will force a spectral gap to close~\citep{PhysRevLett.113.046802}, or topological transitions occur due to disappearance of energy gaps.
Our finding may help understanding better the nature of disorder-induced topological transitions.

\section{Conclusion}
\label{Sec:conclusion}

We have studied the effect of disorder on topological properties of zigzag arrays of nanoparticles with polarization-dependent interaction.
For equal-spacing arrays, even in the presence of angle disorder, we have found that the system will remain in the topological regime supporting edge states, and this feature was found to be in agreement with experiments for finite-extend arrays of silicon nanoparticles with an introduced disorder.
For staggered-spacing arrays, we have found richer topological phases and observed that the angle disorder may induce topological phase transitions.
Remarkably, in this latter case the spectral gap remains open even for a strong disorder, and we believe this system provides the first example of a topologically nontrivial system with disordered parameters.

\begin{acknowledgments}
This work has been supported by the Key-Area Research and Development Program of Quandong Province under Grants No. 2019B030330001, the National Natural Science Foundation of China (NNSFC) under Grants No. 11874434 and No.11574405, and the Science and Technology Program of Guangzhou (China) under Grant No. 201904020024. Research of Y.K. is supported by the Office of China Postdoctoral Council (Grant No. 20180052), the National Natural Science Foundation of China (Grant No. 11904419), and the Australian Research Council (Grant DP200101168).
\end{acknowledgments}

\appendix
\section{Detailed experimental data}
\begin{figure}[htp]
	\includegraphics[width = \columnwidth]{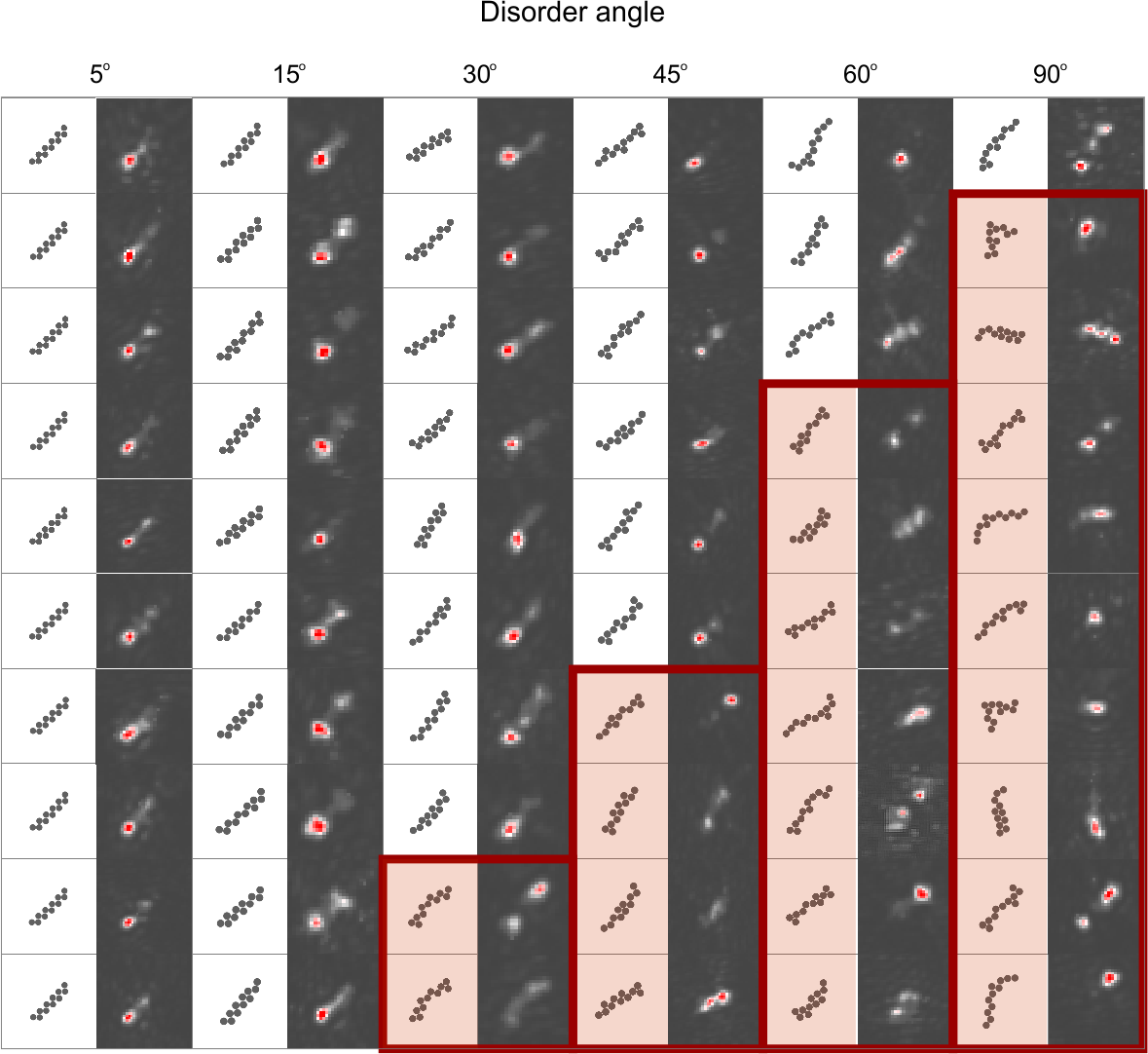}
	\caption{\label{fig:FIG_Exp_Suppl}
		Experimentally observed field distributions in 60 realisations of disordered zig-zag arrays illuminated by linearly polarised light.
		}
\end{figure}

We present in Fig.~\ref{fig:FIG_Exp_Suppl} results on experimentally accumulated statistics of field distributions in zig-zag chains with angular disorder varying from $\delta\psi=5^\circ$ to $\delta\psi=90^\circ$. All chains are 11 disks long, and their configurations are schematically shown in left-side columns. The field distributions along the chains are studied via nonlinear imaging: the chains are excited by short-pulse, high peak power laser pulses at 1590 nm wavelength, and generation of third harmonic signal is imaged onto a camera. The corresponding distributions of the third harmonic are shown as false-colour images in the right-side columns. In the chosen experimental configuration, the topological state is expected to occur at the bottom left edge of the chains. Experimental realisations in which the formation of the edge state was not observed are highlighted with red colour. As the level of disorder increases, the localisation length of the edge states increases reaching and further surpassing the length of the 11-disk chains. Large localisation length in highly disordered zig-zags reduces the probability of edge states observation in finite-length chains.

\end{CJK*}
\end{document}